\documentclass[lettersize,journal]{IEEEtran}
\usepackage{amsmath,amsfonts}
\usepackage{amssymb}
\usepackage{algorithm}
\usepackage{algorithmic}
\usepackage{array}
\usepackage[caption=false,font=normalsize,labelfont=sf,textfont=sf]{subfig}
\usepackage{textcomp}
\usepackage{stfloats}
\usepackage{color}
\usepackage{url}
\usepackage{verbatim}
\usepackage{cite}
\usepackage{mathrsfs}
\usepackage{bbding}
\usepackage{fontawesome}
\usepackage{subcaption}
\usepackage{pifont}
\usepackage{graphicx}
\usepackage{subfig}
\usepackage{booktabs}
\usepackage{multirow}

\newtheorem{remark}{\textbf{Remark}} 

\begin{document}

\title{Hybrid Digital and Analog Airy Beamforming for Near-Field Multi-User Communications}

\author{
\IEEEauthorblockN
{
Shupei Zhang, \IEEEmembership{Graduate Student Member, IEEE}, Boya Di, \IEEEmembership{Senior Member, IEEE},\\ and Lingyang Song, \IEEEmembership{Fellow, IEEE}
}
\thanks{Shupei Zhang, Boya Di and Lingyang Song are with State Key Laboratory of Photonics and Communications, School of Electronics, Peking University, Beijing, China (emails: zhangshupei@pku.edu.cn; diboya@pku.edu.cn; lingyang.song@pku.edu.cn).
}
}
\maketitle

\begin{abstract}
The demands for high data rates in 6G networks have driven the transition toward higher frequencies and larger antenna apertures, giving rise to the near-field communications. In the near-field region, spherical waves enable beam focusing to enhance the received power. However, high-frequency focused beams are highly susceptible to ubiquitous obstacles due to rectilinear trajectories. Particularly in multi-user communications with hybrid precoding, focused beamforming suffers from impaired spectral efficiency under potential multi-user link blockages.
In this paper, we propose an Airy beamforming enabled multi-user transmission scheme. The near-field Airy wavefront with a bending trajectory is first developed to cope with the obstructed channels, possessing the dual capability of bypassing obstacles and concentrating energy. Moreover, a low-complexity Airy beamforming enabled multi-user communication scheme is designed. Specifically, Airy beams capable of circumventing obstacles and aligning with users are first obtained through hierarchical Airy beam training. Then, the selected Airy beams are leveraged to configure the analog beamformer to achieve multi-user obstacle-avoiding access without full channel state information acquisition.
Finally, the digital beamformer is utilized to further mitigate inter-user interference.
In simulations, the beam patterns demonstrate that the proposed Airy beamforming successfully circumvents blockages and aligns with multiple users. Across typical mmWave to THz bands, the proposed scheme outperforms conventional focused beamforming in terms of spectral efficiency.
\end{abstract}

\begin{IEEEkeywords}
Airy beam, multi-user, hybrid beamforming, near-field.
\end{IEEEkeywords}

\section{Introduction}

To satisfy the escalating demands for high data rates and ubiquitous connectivity in 6G networks, higher carrier frequencies and larger antenna apertures are employed to enhance spectral efficiency~\cite{6G1,6G2}. As the frequency and aperture size continue to scale, the near-field region of the transceiver expands significantly~\cite{6G3,6G4}. Unlike the linear phase response of far-field electromagnetic~(EM) waves, the non-linear phase response of near-field spherical waves enables \emph{beam focusing}~\cite{Focus1}. This allows EM energy to be concentrated at a specific focal point rather than just along a certain direction, thereby substantially improving the received power for users~\cite{Focus2}.

Despite the prominent gain provided by beam focusing, high-frequency focused beams are inherently vulnerable to dynamic environments, particularly the ubiquitous obstacles in the near-field region~\cite{Obs1}. Specifically, high-frequency signals suffer from significant penetration losses. As frequencies rise and wavelengths shorten, EM energy is more readily absorbed when passing through various media, leading to an increase in signal dead zones~\cite{Obs2}. What's more, focused beams converge toward the target along rectilinear trajectories, even minor obstructions may disrupt the focal spot. Consequently, the performance of focused beamforming is compromised in obstructed environments, leading to a precipitous drop in the spectral efficiency.
In near-field multi-user communications, the issue of link blockage becomes even more critical and complex. If conventional focused beamforming is employed, the links of multiple users may be simultaneously blocked, leading to a drastic drop in the overall spectral efficiency.

To cope with the obstructed environments, some initial works have explored new wavefronts to bypass obstacles~\cite{Wave1,Wave2}. Among these, the \emph{Airy beam} is a unique wavepacket characterized by its self-accelerating property~\cite{Airy1}. In contrast to the rectilinear propagation of traditional focused beams, the curved trajectory of Airy beams offers the potential to circumvent obstacles~\cite{Airy2}.
In the single-user case, the base station~(BS) directly selects an appropriate Airy beam whose curved trajectory avoids obstacles while aligning with the user, thereby establishing a reliable communication link in obstructed environments.

Unlike single-user communications, extending Airy beam based communications to multi-user cases encounters new issues that need to be addressed.
Subject to limited communication resources such as power consumption and RF chains, the BS needs to simultaneously generate multiple Airy beams, each tailored to bypass obstacles and align with a specific user.
Meanwhile, beyond obstacle avoidance, the concurrent transmission to multiple users necessitates interference mitigation to further enhance overall system spectral efficiency.


Existing works mainly focus on multi-user near-field communications based on focused beamforming~\cite{Multi1,Multi2,Multi3,Multi4,Multi5}, or single-user communications aided by Airy beams~\cite{airy1,airy2,airy3,airy4}.
Authors in~\cite{Multi1} propose near-field location division multiple access, where the polar codebook is utilized to design the hybrid beamformer.
Near-field wideband multi-user communication is investigated in~\cite{Multi2} to mitigate the spatial wideband effect of beam focusing.
Work \cite{Multi3} leverages a small number of far-field wide beams and employs graph neural networks to map the optimal focused beams for multiple users.
Leveraging the Fourier spectrum of the Airy beam, the authors in~\cite{airy1} incorporate a curvature phase into the focused beam to generate curved beams with a bending trajectory.
To identify the optimal curved beam for the user,~\cite{airy2} proposes a two-stage beam training scheme to mitigate the search overhead.
In~\cite{airy3}, the physical characteristics and link performance of Airy beams in free-space and partially obstructed cases are analyzed, demonstrating the performance advantages of Airy beams in obstructed communications.

However, existing schemes are not directly applicable to multi-user communications in obstructed environments. In this paper, we consider Airy beamforming aided multi-user communications. To mitigate the hardware cost and power consumption of fully digital beamformers, a hybrid beamforming architecture is adopted~\cite{Hyb1,Hyb2}. Nevertheless, unlike conventional hybrid beamforming for multi-user communications, applying Airy beamforming is challenging due to the following reasons.
\emph{First}, channel estimation for obstructed links is only theoretically feasible. Consequently, it remains non-trivial to identify the optimal Airy beams for all users in the absence of channel state information~(CSI).
\emph{Second}, the Airy beams with obstacle-avoiding trajectories are jointly generated by the analog and digital beamformers. However, maintaining these specific trajectories for all users while simultaneously eliminating inter-user interference within the hybrid beamforming is intricate.

By addressing these challenges, this paper proposes an Airy beamforming enabled multi-user communication scheme with the hybrid beamformer. Tailored to the characteristics of near-field obstructed channels, we first present a near-field Airy beam wavefront that possesses the dual capability of bypassing obstacles and concentrating energy.
Leveraging this wavefront, the Airy beam codebook and multi-user hierarchical beam training scheme are designed to identify the optimal Airy beams for all users without relying on full CSI.
The selected Airy beams are then employed to configure the high-dimensional analog beamformer, achieving multi-user obstacle-bypassing access and precise alignment. Subsequently, only the low-dimensional effective channel needs to be acquired to design the digital beamformer for the interference suppression.
The main contributions of this paper are summarized as follows.
\begin{itemize}
\item To enable beams to bypass obstacles and align with multiple users, we propose a new wavefront named the near-field Airy beam for obstructed channels, which is achieved through the joint design of non-uniform amplitude and non-linear phase responses. Moreover, the method for the hybrid beamformer to generate this wavefront is developed.

\item Leveraging these near-field Airy beams, a low-complexity multi-user transmission scheme for obstructed environments is presented. Specifically, to configure the analog beamformer, the optimal Airy beams capable of circumventing blockages and aligning with users are obtained through hierarchical beam training. Subsequently, only the low-dimensional effective channel needs to be acquired to design the digital beamformer for inter-user interference cancellation.

\item Simulation results demonstrate that the beam patterns of the proposed Airy beamforming successfully bypass obstacles and align with multiple users. It is shown that the proposed scheme outperforms conventional focused and steered beamforming across typical mmWave (90 GHz) to THz (150 GHz) bands.
\end{itemize}

\begin{figure}[t]
\setlength{\abovecaptionskip}{0pt}
\setlength{\belowcaptionskip}{0pt}
	\centering
    \includegraphics[width=1\columnwidth]{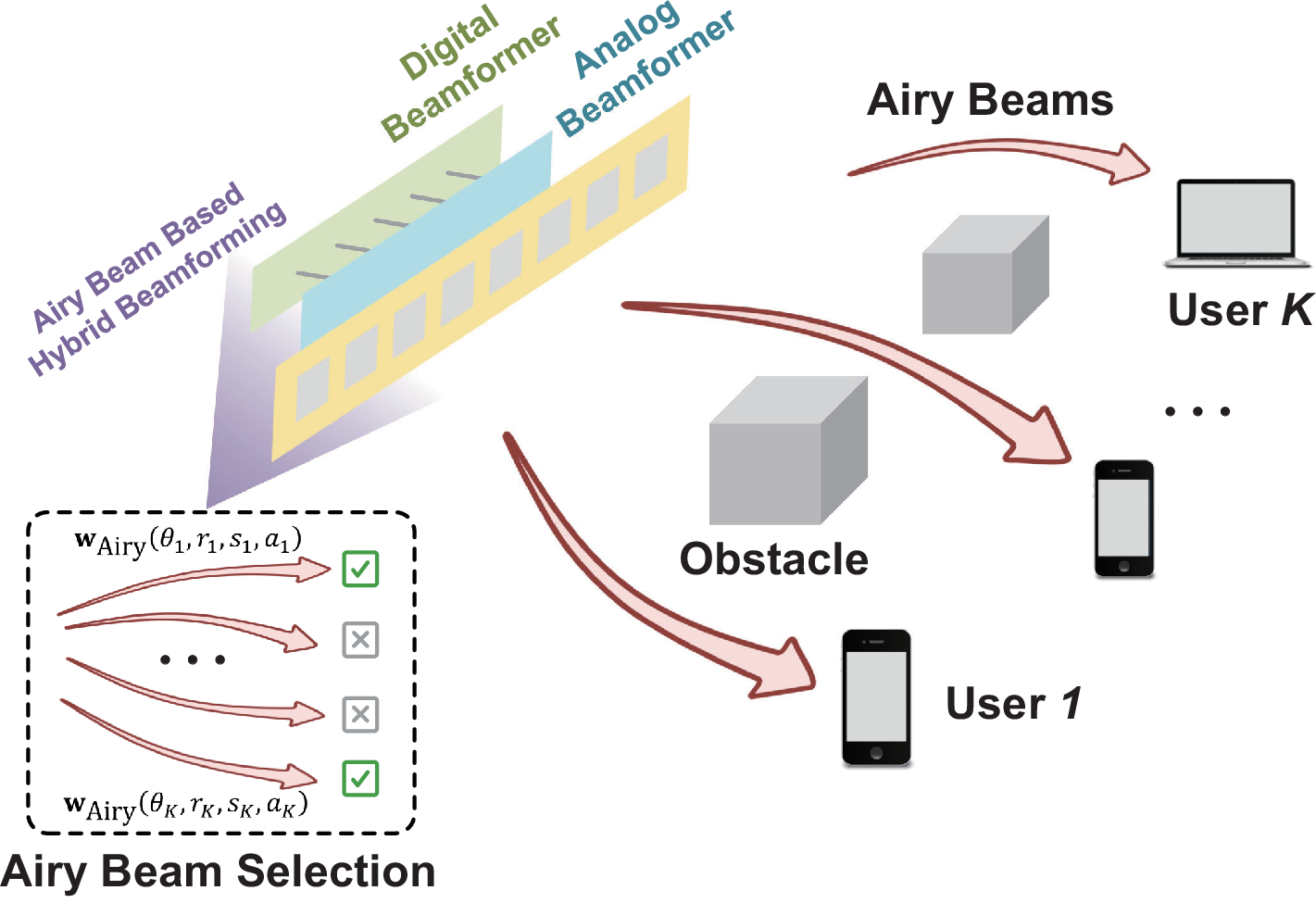}
	\caption{Block diagram of the Airy beam based hybrid beamforming for near-field multi-user communications.}
	\label{Fig:block}
\end{figure}

\emph{Organization}: The remainder of this paper is organized as follows. Section~\ref{Sec-II} presents the system model, including the scenario, EM wave propagation under blockage, and the signal model for multi-user systems. Section~\ref{Sec-III} characterizes the generation of near-ﬁeld Airy beams and analyzes their wavefront matching and performance in obstructed environments. Section~\ref{Sec-IV} presents the design of the Airy beamforming enabled multi-user transmission scheme, including the codebook design, multi-user hierarchical beam search, and the beamformer optimization. Finally, simulation results and conclusions are presented in Section~\ref{Sec-V} and Section~\ref{Sec-VI}, respectively.

\emph{Notations}: Bold lowercase and uppercase letters denote vectors and matrices, respectively. The conjugate transpose and transpose of a matrix are denoted by $(\cdot)^H$ and $(\cdot)^T$, respectively. $\|\cdot\|_2$ and $\|\cdot\|_F$ represent the $l_2$-norm and the Frobenius norm, respectively. $\mathbf{I}_K$ denotes the $K \times K$ identity matrix. $|[\mathbf{A}]_{n,m}|$ and $\angle([\mathbf{A}]_{n,m})$ represent the amplitude and phase of the $(n,m)$-th element of matrix $\mathbf{A}$.

\section{System Model}\label{Sec-II}

This section presents the communication scenario, EM propagation under blockage, and the problem formulation.

\subsection{Scenario Description}

We consider a large-scale array assisted multi-user near-field communication system. As illustrated in Fig.~\ref{Fig:block}, the base station (BS) is equipped with a uniform linear array consisting of $N$ antenna elements to serve $K$ single-antenna users. To achieve a trade-off between hardware cost and spectral efficiency, a hybrid beamforming architecture is adopted. $N$ antenna elements are connected to $N_{RF}$ RF chains, where $K \le N_{RF} \ll N$.
The transmitted signal is processed by a digital beamformer $\mathbf{F}_{\text{D}} \in \mathbb{C}^{N_{RF} \times K}$ and an analog beamformer $\mathbf{F}_{\text{A}} \in \mathbb{C}^{N \times N_{RF}}$.
The ULA is centered at the origin and aligned along the $y$-axis. The coordinates of the $n$-th antenna element are given by $(x_0, y_n) = (0, nd)$, where $d$ is the element spacing and $n \in \{-\frac{N-1}{2}, \dots, \frac{N-1}{2}\}$. The location of the $k$-th user is represented by $(x_k, y_k)$.

In the near-field region, high-frequency signals may suffer from severe penetration losses. Consequently, ubiquitous obstacles can easily block the direct line-of-sight (LoS) links between specific antenna elements and the users, leading to link outages. Hence, it is essential to develop transmission schemes for multi-user communication under blockage.

\subsection{Obstructed EM Wave Propagation}

Traditional geometric channel models are inadequate for capturing diffraction effects in blocked environments~\cite{channel0,channel1}. Consequently, we utilize the Rayleigh-Sommerfeld diffraction theory and the Huygens-Fresnel principle to model the EM wave propagation under obstruction. Let $s=[s_{1},s_{2},...,s_{K}]^{T}\in\mathbb{C}^{K\times1}$ represent the transmitted data symbols for the $K$ users, which satisfy the normalized power condition $\mathbb{E}[\mathbf{s}\mathbf{s}^H] = \mathbf{I}_K$. The BS applies a hybrid beamforming architecture, yielding an equivalent precoding vector $\mathbf{w} = \mathbf{F}_{\text{A}}\mathbf{F}_{\text{D}}\mathbf{s}$. Accordingly, the initial electric field at the $n$-th antenna element can be defined as $E(x_0, y_n) = {w}_n$.

To obtain the electric field $E(x_{k},y_{k})$ at the $k$-th user's location, a step-by-step spatial evolution method is utilized. The propagation plane is discretized into successive parallel planes $x_{i}=(i-1)\Delta x$ (where $i=2,3,...,I$), and the field is iteratively updated from the transmit aperture $(x_{0}=0)$ all the way to the target distance $x_{k}$. A spatial masking function $B(x,y)$ is incorporated to model the physical blockages, taking the value of $0$ for coordinates $(x,y)$ inside an obstacle and $1$ for outside points. During the $i$-th propagation step, the electric field arriving at plane $x_{i}$ is synthesized by integrating the secondary wavelets emitted from the previous plane $x_{i-1}$, yielding~\cite{channel1,channel2,channel3}
$$E(x_{i},y)=B(x_{i},y)\int E(x_{i-1},y^{\prime})\frac{e^{-j\kappa r_{u}}}{2\pi r_{u}^{2}}\Delta x\left(j\kappa+\frac{1}{r_{u}}\right)dy^{\prime}, $$
where $\kappa=2\pi/\lambda$ represents the wavenumber, and $r_{u}=\sqrt{(y-y^{\prime})^{2}+\Delta x^{2}}$ defines the distance from the source point $(x_{i-1},y^{\prime})$ to the observation point $(x_{i},y)$. 

The angular spectrum method~(ASM) can be used to simplify the calculation of the above integral. By carrying out this progressive calculation up to the $k$-th user's plane $x_{k}$, a linear mapping between the transmit beamformer $\mathbf{w}$ and the final electric field $E(x_{k},y_{k})$ is constructed, expressed as\begin{equation}
E(x_k, y_k) = \mathbf{h}_k^H \mathbf{w},
\end{equation}
where $\mathbf{h}_k \in\mathbb{C}^{N\times1}$ is the equivalent channel vector for the $k$-th user, inherently capturing the complex diffraction effects induced by the obstacles.

\subsection{Signal Model and Problem Formulation}

The received signal for the $k$-th user, after the EM wave propagates through the obstructed environment, can be represented as 
\begin{equation}
y_k = \mathbf{h}_k^H \mathbf{F}_{\text{A}} \mathbf{f}_{\text{D},k} s_k + \sum_{j \neq k}\mathbf{h}_k^H \mathbf{F}_{\text{A}} \mathbf{f}_{\text{D},j} s_j + n_k,
\end{equation}
where $\mathbf{f}_{\text{D},k}$ is the $k$-th column of the digital beamformer and $n_k \sim \mathcal{CN}(0, \sigma^2)$ is the additive white Gaussian noise.
The achievable data rate for the $k$-th user is then formulated as
\begin{equation}
R_k = \log_2 \left( 1 + \frac{|\mathbf{h}_k^H \mathbf{F}_{\text{A}} \mathbf{f}_{\text{D},k}|^2}{\sum_{j \neq k} |\mathbf{h}_k^H \mathbf{F}_{\text{A}} \mathbf{f}_{\text{D},j}|^2 + \sigma^2} \right).
\end{equation}
The sum rate maximization problem is then formulated as\begin{subequations}\begin{align}\max_{\mathbf{F}_{\text{A}}, \mathbf{F}_{\text{D}}} \quad & \sum_{k=1}^{K} R_k \\
\text{s.t.} \quad & \|\mathbf{F}_{\text{A}} \mathbf{F}_{\text{D}}\|_F^2 \le P, \\
& \left|[\mathbf{F}_{\text{A}}]_{n,m}\right| = \frac{1}{\sqrt{N}}, \quad \forall n, m, \label{con2}
\end{align}
\end{subequations}
where $P$ denotes the total transmit power. Constraint~\eqref{con2} ensures the constant modulus of the analog phase shifters.

Due to the large scale of the antenna array and the ubiquitous presence of obstacles in the near-field region, estimating the complete full-dimensional CSI for all users involves prohibitive overhead.  Therefore, it is necessary to design an efficient multi-user transmission scheme tailored for such cases.
Considering the high dimensionality of the analog beamformer, we adopt codebook-based beam training to configure $\mathbf{F}_{\text{A}}$ without full CSI. Once the analog beamformer is determined, the BS only needs to estimate the low-dimensional equivalent channel, given by $\mathbf{H}_{\text{eq}} = \mathbf{H}^H \mathbf{F}_{\text{A}}$ where $\mathbf{H} = [\mathbf{h}_1, \dots, \mathbf{h}_K] \in \mathbb{C}^{N \times K}$, to design the digital beamformer $\mathbf{F}_{\text{D}}$ for interference mitigation.

However, the conventional near-field codebooks are designed for beam focusing under LoS conditions. It may not comprehensively account
for the intricate characteristics of obstructed near-field channels. This necessitates the development of new wavefronts and specialized codebook designs for obstructed communications.

\section{Wavefront Design of Near-Field Airy Beams}\label{Sec-III}

In this section, we propose the near-field Airy beam and evaluate its performance in obstructed environments.

\subsection{Guidelines of Wavefront Design Under Blockage}\label{Sec_Guideline}

In obstructed near-field communications, exploring wavefronts with curved trajectories to bypass obstacles appears to be an intuitive solution. However, to establish the beamforming design principle, we first analyze the limitations of the received gain under blockages.
Let $\mathbf{h}_k$ denote the equivalent channel response vector from the array to the $k$-th user, where its $n$-th element has an amplitude $\eta_{k,n} = |h_{k,n}|$. For an arbitrary transmit beamforming vector $\mathbf{w} \in \mathbb{C}^{N \times 1}$ subject to a total power constraint $\|\mathbf{w}\|_2^2 \le P$, the received gain $G_k$ of the user is given by $G_k = |\mathbf{h}_k^H \mathbf{w}|^2$. According to the Cauchy-Schwarz inequality, this received gain is bounded by$$G_k = |\mathbf{h}_k^H \mathbf{w}|^2 \le \|\mathbf{h}_k\|_2^2 \|\mathbf{w}\|_2^2 = P \sum_{n=1}^N \eta_{k,n}^2.$$This formulation reveals that the upper bound of the received gain is inherently determined by the channel's amplitude norm $\|\mathbf{h}_k\|_2^2$.
Based on this bound, we first consider the case of complete blockage, where a user is located in a deep shadow region and all direct LoS paths are intercepted. In this case, the channel relies entirely on NLoS edge diffracted paths. According to the geometrical theory of diffraction, at high frequencies such as mmWave and THz bands, the diffraction coefficient plummets, leading to severe attenuation where $\eta_{k,n} \to 0$ for all $n$~\cite{diffraction}. This proves that in completely obstructed regions, the performance is fundamentally bottlenecked by the channel attenuation. Under such channel conditions, solely relying on wavefront design cannot improve the gain, regardless of the trajectory curvature.

\begin{remark}\label{remark1}
When the LoS paths from all antenna elements to the user are blocked, the signal cannot be delivered to the user directly. Constrained by the severe degradation of the channel link gain, hardly any wavefront design can effectively improve the received power.
\end{remark}

Therefore, the focus of beamforming design pivots to partially obstructed cases. In these blocked cases, obstacles sever a part of the LoS paths, leaving the surviving unblocked paths as the primary avenues for energy transfer. This partial blockage induces a highly non-uniform amplitude distribution across the antenna array, where unblocked elements retain dominant gains while blocked ones suffer near-zero attenuation.
Returning to the Cauchy-Schwarz inequality, under the transmit power constraint, the equality holds if $\mathbf{w} \propto \mathbf{h}_k$. This dictates that the effective beamformer should explicitly match the channel response. Rather than attempting to synthesize a heavily curved trajectory, the beamformer should exploit the surviving direct paths by aligning with the non-uniform amplitude distribution of the obstructed channel. Motivated by this, we propose the near-field Airy beam in the following subsection, a wavefront specifically formulated to match the characteristics of partially obstructed channels through its amplitude and phase manipulation.

\subsection{Generation of Near-Field Airy Beams}\label{III-A}

\begin{figure}[t]
\setlength{\abovecaptionskip}{0pt}
\setlength{\belowcaptionskip}{0pt}
	\centering
    \includegraphics[width=1\columnwidth]{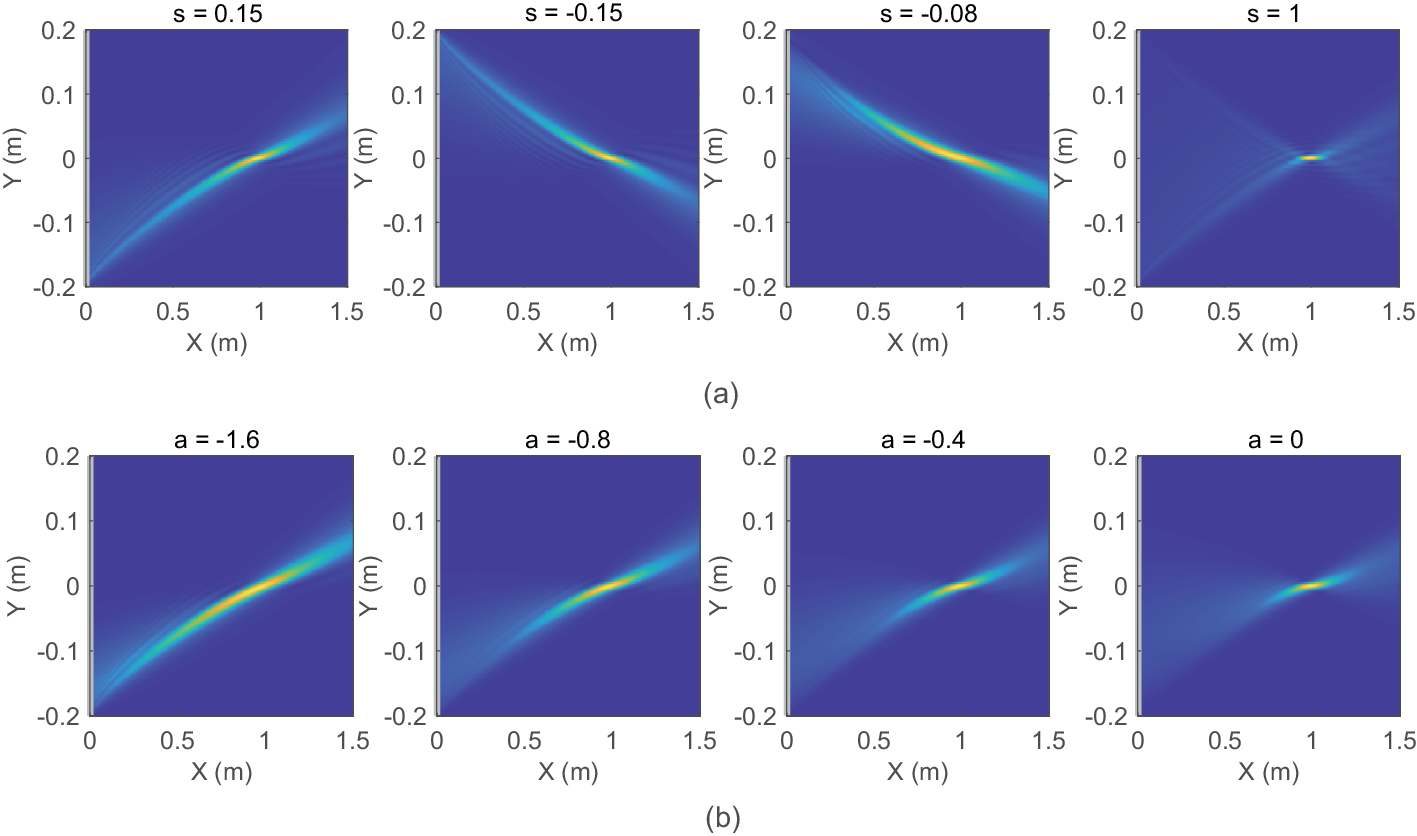}
	\caption{Impact of parameters (a) spatial scaling factor $s$ and (b) exponential decay parameter $a$ on Airy beam patterns.}
	\label{Fig:pattern}
\end{figure}

In unobstructed environments, to match the channel response, the focused beam employs a near-field non-linear phase response and a uniform amplitude response. However, in obstructed environments, the uniform amplitude response becomes inadequate due to the presence of blockages. Similarly, the curved beam primarily focuses on generating a bending trajectory through phase regulation while adopting the same uniform amplitude response as the focused beam. Unlike conventional wavepackets, classical Airy beams employ a non-uniform amplitude response and possess unique properties such as being non-diffracting and self-accelerating.
The response of the $n$-th antenna element for a classical Airy beam can be expressed as~\cite{airy1}
\begin{equation}
\tilde{w}_{\text{Airy}, n}(s,a) = {Ai}\left( \frac{nd}{s} \right) \exp\left( a \frac{nd}{s} \right),
\end{equation}
where $nd$ is the coordinate of the $n$-th element, $\text{Ai}(\cdot)$ denotes the Airy function, $s$ is the spatial scaling factor, and $a$ is the exponential decay parameter.

To overcome the performance bottlenecks of rectilinear focused beams in obstructed environments, we propose the near-field Airy beam. This new wavefront is synthesized by superimposing the focusing phase onto the amplitude envelope of classical Airy wavepackets. Unlike conventional solutions, this joint design allows the beam to exploit the non-linear phase characteristics of the near field while matching the non-uniform amplitude responses of obstructed channels~\cite{focus_airy}. The response for the $n$-th antenna element of a fully digital array is formulated as
\begin{equation}
w_{\text{Airy}, n}(\theta, r, s, a) = \frac{1}{\mathcal{N}} \underbrace{\tilde{w}_{\text{Airy}, n}}_{\text{Amplitude}} \cdot \exp(j \underbrace{\Phi_n(r, \theta)}_{\text{Phase}}),
\end{equation}
where ${\mathcal{N}}$ is the normalization factor ensuring $\|\mathbf{w}\|_2^2 = P$, and $\Phi_n(r, \theta)$ represents the focusing phase profile, given by $\Phi_n(r, \theta) = \frac{2\pi}{\lambda}(- nd \sin\theta +  \frac{n^2d^2\cos^2\theta}{2r})$.

The geometric and energetic characteristics of this wavefront are governed by four parameters $(\theta,r,s,a)$.
\begin{itemize}
\item  Focal Point $(\theta, r)$: These parameters specify the angular direction and the focusing distance, ensuring the EM energy is concentrated at the intended user location.
\item Trajectory Curvature $(s)$: The spatial scaling factor $s$ determines the orientation and intensity of the beam's self-bending path. This allows the main lobe to follow a curved trajectory that circumvents blockages.
\item Energy Distribution $(a)$: The exponential decay parameter $a$ regulates the width and energy density of the main lobe.
\end{itemize}
As shown in Fig.~\ref{Fig:pattern}, the spatial scaling factor $s$ determines the direction and degree of the beam curvature to bypass obstacles. Meanwhile, the exponential decay parameter $a$ controls the energy distribution of the main lobe.
As $|s|$ increases and $|a|$ decreases, the near-field Airy beam gradually degenerates into the conventional focused beam.

\subsection{Wavefront Matching with Obstructed Channels}

Compared to conventional focused beams that rely solely on phase manipulation for energy concentration, the proposed near-field Airy beam introduces an additional degree of freedom in amplitude regulation. Based on the Airy function, this amplitude regulation not only facilitates the generation of the curved main lobe trajectory to bypass obstacles but also provides the capability to flexibly adapt to the amplitude response of obstructed channels.

\begin{figure}[t]
\setlength{\abovecaptionskip}{0pt}
\setlength{\belowcaptionskip}{0pt}
	\centering
    \includegraphics[width=1\columnwidth]{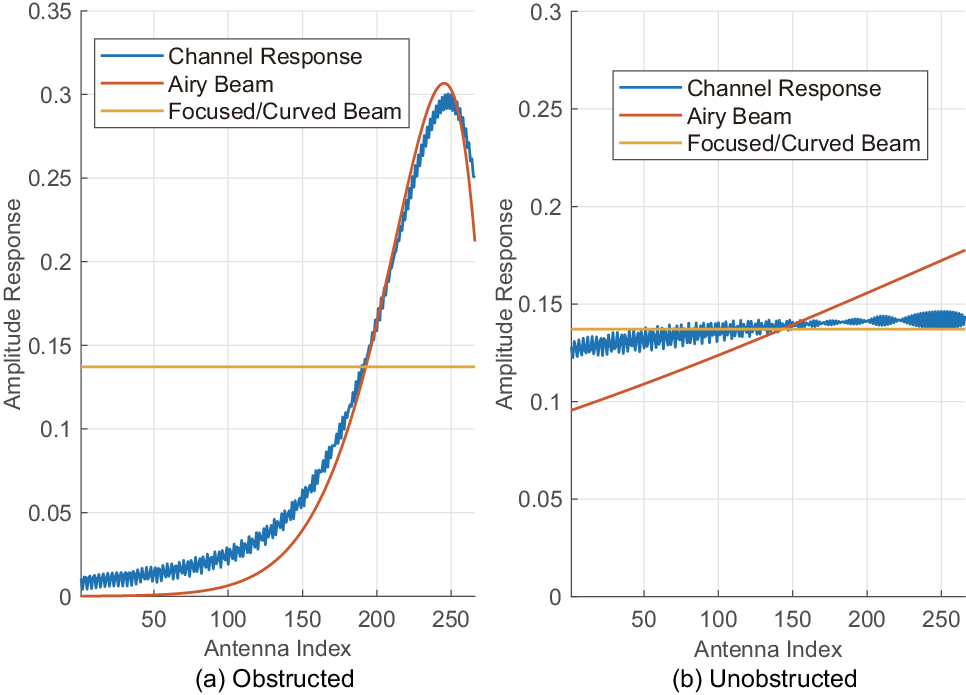}
	\caption{Amplitude matching between wavefronts and channels under (a) obstructed and (b) unobstructed cases.}
	\label{Fig:amp_response}
\end{figure}

To characterize this, we decompose the equivalent channel of the $k$-th user into its amplitude and phase components as $h_{k,n} = \eta_{k,n} e^{-j\Psi_{k,n}}$, where $\eta_{k,n}$ and $\Psi_{k,n}$ represent the channel amplitude and phase, respectively.
For a given beamforming vector $\mathbf{w}$ with $w_n = a_n e^{j\Phi_n}$, we assume the beamforming phase can be aligned with the channel phase, i.e., $\Phi_n = \Psi_{k,n}$. This is because the unobstructed links can still leverage the near-field focusing phase to achieve phase alignment.
Accordingly, the received gain can be expressed as$$G_k = \left| \sum_{n=1}^{N} h_{k,n}^* w_n \right|^2 = \left( \sum_{n=1}^{N} \eta_{k,n} a_n \right)^2.$$
It implies that the user's received power mainly depends on the degree of alignment between the beam's amplitude distribution $a_n$ and the channel's amplitude response $\eta_{k,n}$. Based on this, two typical cases are discussed.

\textbf{Obstructed Case:} In the presence of blockages, the obstacles sever the direct links between specific antenna elements and the user. Consequently, the link gains corresponding to the blocked antenna elements experience a precipitous drop, resulting in a highly non-uniform amplitude distribution. As illustrated in Fig.~\ref{Fig:amp_response}~(a), the near-field Airy beam exploits its non-uniform amplitude regulation to effectively match this irregular channel amplitude response, thereby enhancing the user's received gain.
In contrast, the focused and curved beams, which lack amplitude regulation, fail to match the obstructed channel's amplitude response.

\textbf{Unobstructed Case:} Conversely, in unobstructed environments, the LoS paths are preserved, and the link gains from the user to all antenna elements remain approximately equal (i.e., uniform $\eta_{k,n}$). Under such conditions, the Airy function flattens its amplitude variation by increasing $|s|$ and decreasing $|a|$. As depicted in Fig.~\ref{Fig:amp_response}~(b), both the near-field Airy beam and the focused/curved beams can approximate the uniform channel amplitude response.

In summary, the proposed near-field Airy beam exhibits applicability across both obstructed and unobstructed cases. Particularly in obstructed environments, its unique capability to adapt to drastic channel amplitude variations makes it effective at mitigating blockage-induced power losses.

\begin{figure}[t]
\setlength{\abovecaptionskip}{0pt}
\setlength{\belowcaptionskip}{0pt}
	\centering
    \includegraphics[width=1\columnwidth]{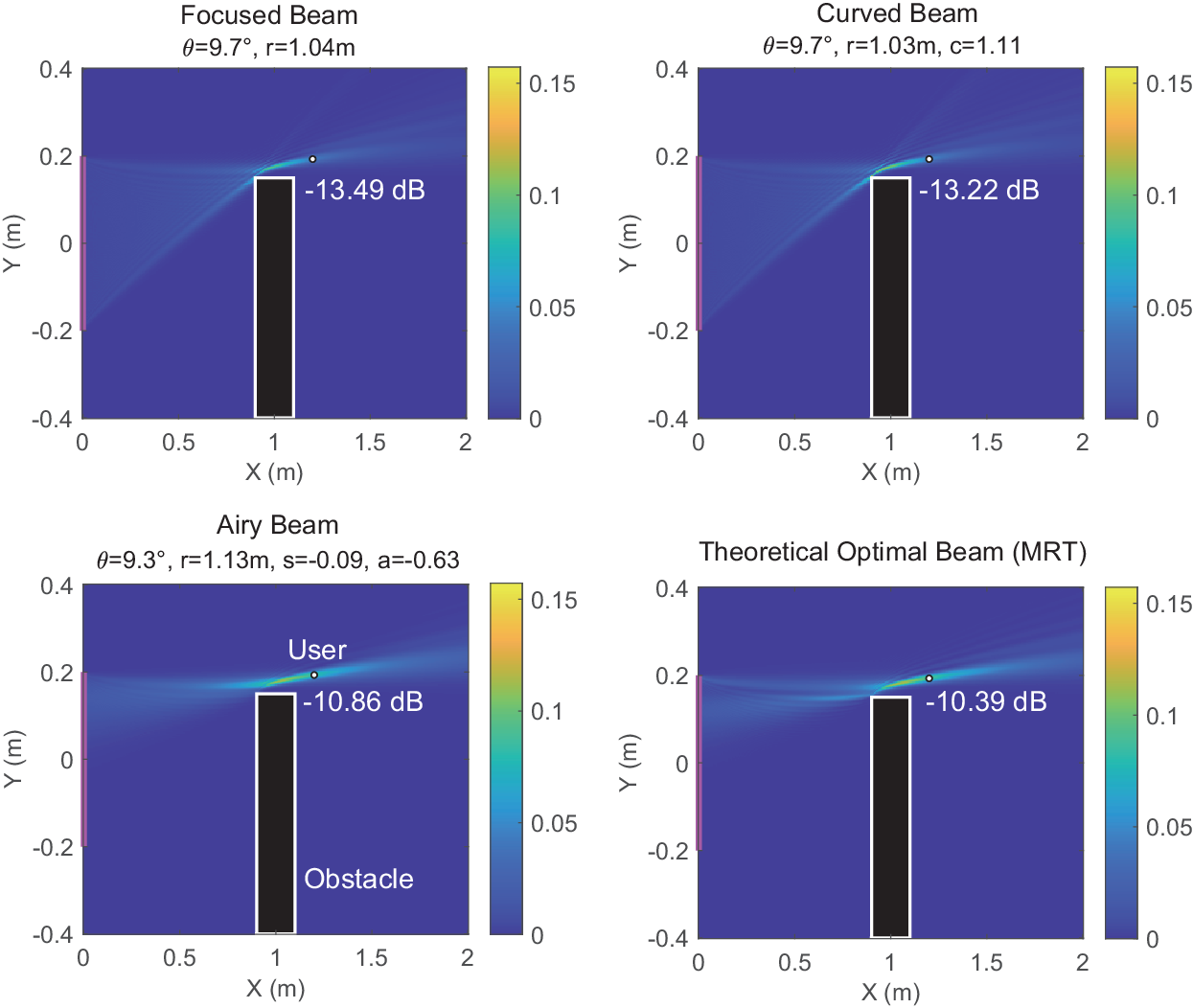}
	\caption{Optimal beam patterns of focused, curved, Airy, and MRT Beams. The frequency is 100GHz, and the number of antennas is 266.}
	\label{Fig:4beam}
\end{figure}

\subsection{Performance of Near-Field Airy Beams in Obstructed Environments}


Traditional near-field beamforming strategies, such as focused and curved wavefronts, rely on phase-only adjustments to achieve focal concentration or trajectory bending. However, this single-dimensional control is often insufficient in severely obstructed environments. The proposed near-field Airy beam addresses this limitation by jointly regulating both the non-linear phase and the amplitude distribution of the EM waves. By integrating the spatially varying amplitude characteristics of the Airy function, this new wavefront matches the non-uniform gain profile of the blocked channel, achieving a curved path capable of bypassing obstructions.

To compare the performance of different wavefronts in obstructed environments,
Fig.~\ref{Fig:4beam} illustrates the optimal beam patterns for various wavefronts under blockage.
The obstacle is located at $x \in [0.9, 1.1]$ m and $y \in [-0.4, 0.15]$ m, and the user is positioned at $(1.2, 0.19)$ m. Optimal configurations for the focused, curved, and near-field Airy wavefronts are determined via an exhaustive parameter search.
The maximum ratio transmission~(MRT) beam is designed assuming perfect CSI, serving as a performance upper bound.
It is observed that the focused beam fails to concentrate energy at the user location due to the blockage. While the curved beam maintains a bending trajectory, its resulting gain at the user is limited. Conversely, the near-field Airy beam effectively bypasses the obstruction while concentrating its main lobe on the target, yielding a received power nearly identical to the MRT upper bound.
The beam pattern of the optimal near-field Airy beam closely approximates that of the MRT beam, which validates the wavefront design guidelines in Section~\ref{Sec_Guideline}.
It confirms that the received power in obstructed environments is primarily enhanced by effectively leveraging the surviving LoS paths.

\section{Airy Beamforming Enabled Obstructed Multi-User Communications}
\label{Sec-IV}

In this section, the procedure of Airy beamforming enabled multi-user communications is described, and the specific beamforming strategy is developed.

\subsection{Airy Beam Codebook for Hybrid Beamforming}\label{III-B}

\begin{figure*}[t]
\setlength{\abovecaptionskip}{0pt}
\setlength{\belowcaptionskip}{0pt}
	\centering
    \includegraphics[width=1.95\columnwidth]{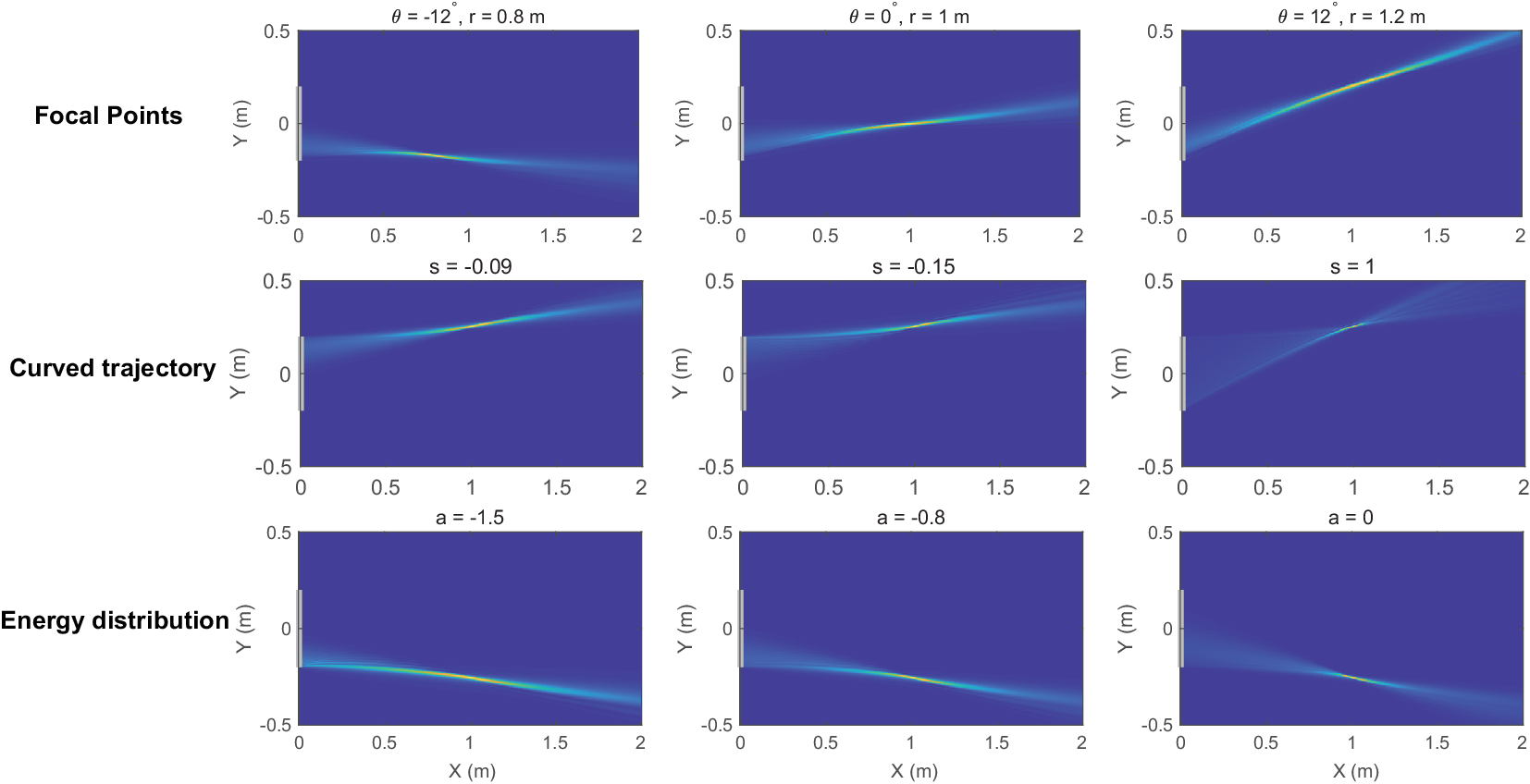}
	\caption{Codewords in the Airy beam codebook: varying focal points, curved trajectories, and energy distributions.}
	\label{Fig:codewords}
\end{figure*}

In practical communications, the exact locations of the users and the obstacles are typically unknown to the BS. Consequently, the optimal near-field Airy beam that aligns with the user cannot be obtained directly. To address this, it is essential to design a predefined near-field Airy beam codebook to search for the optimal beam.
The near-field Airy beam codebook is constructed by discretizing the parameter space $(r, \theta, s, a)$. The codebook set can be defined as
$$\mathcal{W}_{\text{Airy}} = \left\{ \mathbf{w}_{\text{Airy}}(\theta, r, s, a) \mid \theta \in \Theta, r \in \mathcal{R}, s \in \mathcal{S}, a \in \mathcal{A} \right\},$$
where $\Theta = \{ \theta_1, \theta_2, \dots, \theta_{N_\theta} \}$, $\mathcal{R} = \{ r_1, r_2, \dots, r_{N_r} \}$, $\mathcal{S} = \{ s_{1}, s_{2}, \dots, s_{N_s} \}$, and $\mathcal{A} = \{ a_1, a_2, \dots, a_{N_a} \}$ represent the discrete sampling sets for the steering angle $\theta$, focusing distance $r$, spatial scaling factor $s$, and exponential decay parameter $a$, respectively. 
Here, $N_\theta, N_r, N_s, \text{ and } N_a$ represent the number of discrete samples for the respective parameters.
Considering the correlation among codewords, the angle $\theta$ and parameter $a$ are sampled uniformly, while the distance $r$ and parameter $s$ are sampled non-uniformly.
As illustrated in Fig.~\ref{Fig:codewords}, through the Airy beam codebook design, the codebook contains near-field Airy beams with varying focal points, curved trajectories, and energy distributions to accommodate diverse obstacle configurations and user locations.

The near-field Airy beam $\mathbf{w}_{\text{Airy}}$ defined in Section~\ref{III-A} assumes a fully digital architecture, where each antenna is controlled by a dedicated RF chain. To realize this wavefront by hybrid beamforming, the desired Airy beam is approached using the analog beamformer $\mathbf{F}_{\text{A}}$ and one column of the digital beamformer $\mathbf{f}_{\text{D}}$.
For a desired airy beam $\mathbf{w}_{\text{T}} = \mathbf{w}_{\text{Airy}}(r, \theta, s, a)$, the optimization problem is formulated as
\begin{equation}
\begin{aligned}~\label{codeword}
\min_{\mathbf{F}_{\text{A}}, \mathbf{f}_{\text{D}}} & \quad \left\| \mathbf{w}_{\text{T}} - \mathbf{F}_{\text{A}} \mathbf{f}_{\text{D}} \right\|_2^2 \\
\text{s.t.} \quad & \|\mathbf{F}_{\text{A}} \mathbf{f}_{\text{D}}\|_F^2 \le P, \\
& \left|[\mathbf{F}_{\text{A}}]_{n,m}\right| = \frac{1}{\sqrt{N}}, \quad \forall n, m.
\end{aligned}
\end{equation}
To solve the approximation problem~\eqref{codeword}, we utilize an orthogonal matching pursuit based algorithm. The core idea is to iteratively select columns for the analog precoder from a DFT dictionary $\mathbf{D} \in \mathbb{C}^{N \times G}$ that exhibit the highest correlation with the current residual of the target Airy beam. The detailed steps for generating the hybrid Airy beam codeword are summarized in Algorithm~\ref{alg:OMP_Airy}.

\begin{algorithm}[t]
\caption{Airy Beam Codeword for Hybrid Beamforming}
\label{alg:OMP_Airy}
\begin{algorithmic}[1]
\renewcommand{\algorithmicrequire}{\textbf{Input:}}
\renewcommand{\algorithmicensure}{\textbf{Output:}}
\REQUIRE Desired Airy beam $\mathbf{w}_{\text{T}}$, dictionary matrix $\mathbf{D}$.
\STATE \textbf{Initialization:} Residual $\mathbf{r}_0 = \mathbf{w}_{\text{T}}$, index set $\mathcal{I} = \emptyset$, $\mathbf{F}_{\text{A}} = []$, iteration counter $t=1$.
\WHILE{$t \le N_{\text{RF}}$}
    \STATE Calculate correlation: $\mathbf{p} = \mathbf{D}^H \mathbf{r}_{t-1}$.
    \STATE Select best index: $k^* = \arg \max_{k} |\mathbf{p}(k)|$.
    \STATE Update index set: $\mathcal{I} \leftarrow \mathcal{I} \cup \{k^*\}$.
    \STATE Update analog precoder: $\mathbf{F}_{\text{A}} = [\mathbf{F}_{\text{A}}, \mathbf{D}(:, k^*)]$.
    \STATE Update digital precoder: 
    $\mathbf{f}_{\text{D}} = (\mathbf{F}_{\text{A}}^H \mathbf{F}_{\text{A}})^{-1} \mathbf{F}_{\text{A}}^H \mathbf{w}_{\text{T}}$.
    \STATE Update residual: $\mathbf{r}_t = \mathbf{w}_{\text{T}}- \mathbf{F}_{\text{A}} \mathbf{f}_{\text{D}}$.
    \STATE $t \leftarrow t + 1$.
\ENDWHILE
\STATE Power normalization: Scale $\mathbf{f}_{\text{D}}$ such that $\|\mathbf{F}_{\text{A}} \mathbf{f}_{\text{D}}\|_F^2 = P$.
\ENSURE Analog precoder $\mathbf{F}_{\text{A}}$, digital precoder $\mathbf{f}_{\text{D}}$.
\end{algorithmic}
\end{algorithm}

\subsection{Airy Beamforming Enabled Multi-User Transmission Procedure}

To facilitate multi-user communications in obstructed near-field environments, we propose an Airy beamforming based multi-user transmission scheme. The specific procedural steps are delineated as follows.

\textbf{1. Multi-user Beam Training:} Leveraging the hybrid beamforming architecture, the BS performs beam training using the predefined near-field Airy beam codebook $\mathcal{W}_{\text{Airy}}$.
Upon receiving the training signals, each user $k$ identifies the optimal codeword index $n_k^*$ that maximizes its received power. Subsequently, these indices are fed back to the BS.

\textbf{2. Analog Beamformer Design:} Based on the feedback, the BS constructs a fully-digital target beamformer $\mathbf{F}_{\text{T}} \in \mathbb{C}^{N \times K}$. This matrix serves as the desired profile that accounts for both near-field focusing and obstacle bypass.
To map the desired target $\mathbf{F}_{\text{T}}$ onto the hybrid architecture, the analog beamformer $\mathbf{F}_{\text{A}}$ is designed by solving a matrix approximation problem.

\textbf{3. Digital Beamformer Design:}
Given the designed analog beamformer $\mathbf{F}_{\text{A}}$, the BS estimates the low-dimensional equivalent channel $\mathbf{H}_{\text{eq}} = \mathbf{H}^H \mathbf{F}_{\text{A}}$.
Finally, the digital precoder $\mathbf{F}_{\text{D}}$  is designed based on $\mathbf{H}_{\text{eq}}$ to perform interference cancellation.

\subsection{Multi-User Hierarchical Airy Beam Search}

In the multi-user transmission scheme, beam training is initially performed for all $K$ users. Compared to conventional focused beams, the near-field Airy beam introduces new parameters. Consequently, an exhaustive codebook search over the entire parameter space would incur an overhead of $N_{\theta} \times N_{r} \times N_{s} \times N_{a}$. Given that the focused phase term and the Airy amplitude envelope exhibit parameter separability, we propose a low-overhead hierarchical Airy beam search scheme tailored for multi-user cases. This scheme consists of three sequential stages.
\begin{itemize}
\item \textbf{Multi-User Angle-Distance Scanning:} 
In this initial phase, the BS transmits training signals using the near-field polar codebook. This procedure is executed simultaneously for all $K$ users. Based on the received signal power, each user $k \in \{1, \dots, K\}$ identifies and feeds back the order of the optimal angle-distance combination $(\theta_{k}^{*}, r_{k}^{*})$. This stage establishes the basic phase focusing required to support the subsequent obstacle-bypassing transmission.

\item \textbf{Curvature Scanning:} 
Once the specific focal coordinates $(\theta_{k}^{*}, r_{k}^{*})$ are determined for each user, the BS proceeds to optimize the trajectory curvature individually for different users. By fixing the exponential decay parameter at $a=0$, the BS leverages the hybrid Airy beam codewords generated via Algorithm~\ref{alg:OMP_Airy} to scan the spatial scaling factor $s$. Each user $k$ evaluates these codewords to feed back its optimal factor $s_{k}^{*}$, effectively tailoring the beam's bending trajectory to circumvent specific blockages.

\item \textbf{Decay Parameter Scanning:}
Having acquired the optimal $(\theta_{k}^{*}, r_{k}^{*}, s_{k}^{*})$, the BS further refines the beam profile by conducting an individualized scan of the exponential decay parameter $a_{k}^{*}$ for each user. This process maximizes the energy concentration of the main lobe at the user while preserving the previously established obstacle-bypassing capability.
\end{itemize}
At each stage, the users feed back the codeword indices corresponding to their maximum received signal strength.
Ultimately, the BS identifies the optimal near-field Airy $\mathbf{w}_{\text{Airy}}(\theta_k^*, r_k^*, s_k^*, a_k^*)$ for each user $k \in \{1, \dots, K\}$.
It is worth noting that during the entire beam training process, the digital beamformer $\mathbf{f}_{\text{D},k}$ for each user is configured using the uniform $\mathbf{f}_{\text{D}}$ obtained from Algorithm~\ref{alg:OMP_Airy}, as inter-user interference cancellation is not required during this stage. Through this hierarchical strategy, the total training overhead is reduced from a multiplicative complexity to an additive one, specifically $N_{\theta} \times N_{r} + K(N_{s} + N_{a})$. This ensures that the search complexity for Airy beams remains efficient and comparable to that of conventional polar-domain codebooks.

\subsection{Analog and Digital Beamformer Design}
\textbf{Analog Beamformer:}
Following the beam training process, the BS acquires the optimal near-field Airy beamforming vectors $\mathbf{w}_{\text{Airy}}(\theta_k^*, r_k^*, s_k^*, a_k^*)$ for all $K$ users. These optimal codewords implicitly encapsulate the relative positional relationships among the obstacles, the BS, and the users. To facilitate multiple access in obstructed environments, we define a fully digital target beamformer as $\mathbf{F}_{\text{T}} = [\mathbf{w}_{\text{Airy}}(\theta_1^*, r_1^*, s_1^*, a_1^*), \dots, \mathbf{w}_{\text{Airy}}(\theta_K^*, r_K^*, s_K^*, a_K^*)] \in \mathbb{C}^{N \times K}$. The hybrid beamformer is designed to approximate $\mathbf{F}_{\text{T}}$, thereby ensuring that the analog beamformer $\mathbf{F}_{\text{A}}$ is effectively aligned with the users while maintaining the desired beam trajectories. This approximation process can be formulated as
\begin{equation}\label{p2}
\begin{aligned}
\min_{\mathbf{F}_{\text{A}}, \mathbf{F}_{\text{D,temp}}} &\|\mathbf{F}_{\text{T}} - \mathbf{F}_{\text{A}} \mathbf{F}_{\text{D,temp}}\|_F^2, \\  \quad \text{s.t. } &|[\mathbf{F}_{\text{A}}]_{n,m}| = \frac{1}{\sqrt{N}},  \quad \forall n, m,
\end{aligned}
\end{equation}
where $\mathbf{F}_{\text{D,temp}}$ denotes a temporary digital beamformer during the iterative optimization.
We employ the alternating minimization algorithm to solve~\eqref{p2}~\cite{hybrid}.
For a fixed $\mathbf{F}_{\text{A}}$, the optimal $\mathbf{F}_{\text{D,temp}}$ is obtained via singular value decomposition~(SVD), expressed as
$$\mathbf{F}_{\text{T}}^H \mathbf{F}_{\text{A}} = \mathbf{U} \mathbf{\Sigma} \mathbf{V}^H,$$
where $\mathbf{U}$ and $\mathbf{V}$ are unitary matrices, and $\mathbf{\Sigma}$ is a rectangular diagonal matrix containing the singular values. $\mathbf{F}_{\text{D,temp}}$ is then updated by selecting the dominant $K$ singular vectors as $\mathbf{F}_{\text{D,temp}} = \mathbf{V}_1 \mathbf{U}^H$,
where $\mathbf{V}_1 = \mathbf{V}(:,1:K)$ consists of the first $K$ columns of $\mathbf{V}$.
With the optimized $\mathbf{F}_{\text{D,temp}}$, the unconstrained solution for the analog precoder is given by $$\hat{\mathbf{F}}_{\text{A}} = \mathbf{F}_{\text{T}} \mathbf{F}_{\text{D,temp}}^H.$$ To satisfy the constant modulus constraint, we project each element of $\hat{\mathbf{F}}_{\text{A}}$ onto the feasible complex circle, expressed as
\begin{equation}[\mathbf{F}_{\text{A}}]_{n,m} = \frac{1}{\sqrt{N}} \exp\left( j \angle \left( [\hat{\mathbf{F}}_{\text{A}}]_{n,m} \right) \right),\end{equation}where $\angle(\cdot)$ denotes the phase extraction operator.
The algorithm starts with a random phase initialization for $\mathbf{F}_{\text{A}}$, and these steps are repeated for a predefined number of iterations.

\textbf{Digital Beamformer:} 
Once the analog beamformer $\mathbf{F}_{\text{A}}$ is obtained, the estimation of the low-dimensional equivalent channel $\mathbf{H}_{\text{eq}} = \mathbf{H}^H \mathbf{F}_{\text{A}} \in \mathbb{C}^{K \times N_{RF}}$ is performed to facilitate the design of the digital beamformer. Given the reduced dimensionality of the equivalent channel, efficient estimation schemes have been investigated~\cite{Multi1,Multi3}, and thus the details are omitted here for brevity. To eliminate multi-user interference, the low-complexity zero-forcing scheme is adopted, which is expressed as
\begin{equation}
\mathbf{F}_{\text{D}} = \mathbf{H}_{\text{eq}}^H (\mathbf{H}_{\text{eq}}\mathbf{H}_{\text{eq}}^H)^{-1}\mathbf{\Lambda},
\end{equation}
where $\mathbf{\Lambda}$ is the power allocation for users to satisfy the power constraint.

\begin{figure}[t]
\setlength{\abovecaptionskip}{0pt}
\setlength{\belowcaptionskip}{0pt}
	\centering
    \includegraphics[width=0.95\linewidth]{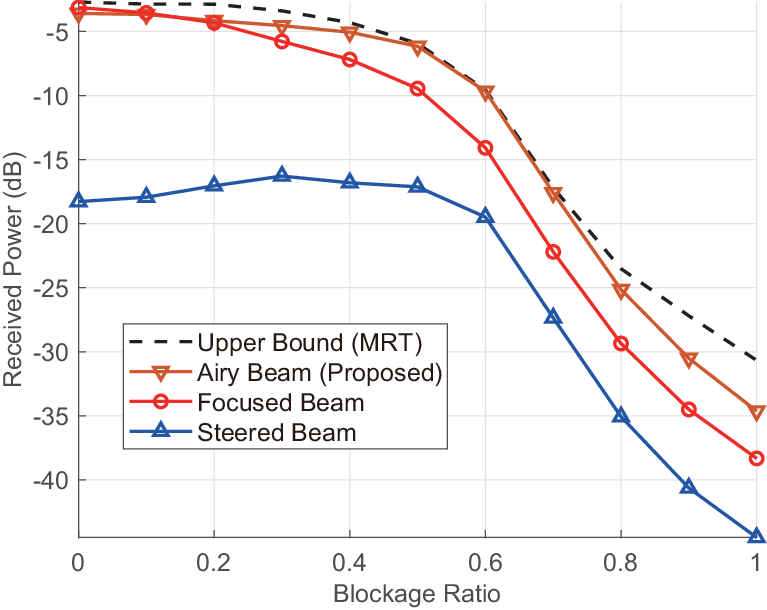}
	\caption{Received power rate vs. blockage ratio.}
	\label{Fig:block_ratio}
\end{figure}

\begin{figure}[t]
\setlength{\abovecaptionskip}{0pt}
\setlength{\belowcaptionskip}{0pt}
	\centering
    \includegraphics[width=0.92\linewidth]{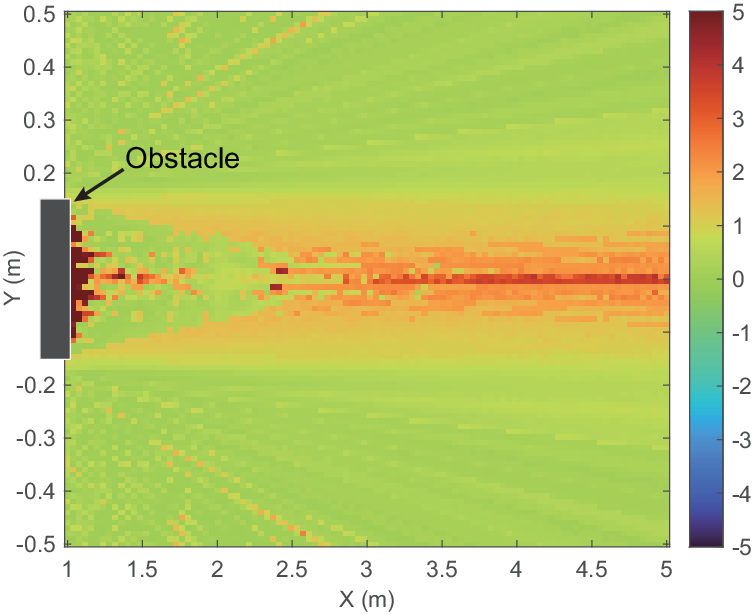}
	\caption{Gain improvement (dB) of exhaustive over hierarchical Airy beam training.}
	\label{Fig:exh_hie}
\end{figure}

\begin{figure}[t]
\setlength{\abovecaptionskip}{0pt}
\setlength{\belowcaptionskip}{0pt}
	\centering
    \includegraphics[width=0.92\linewidth]{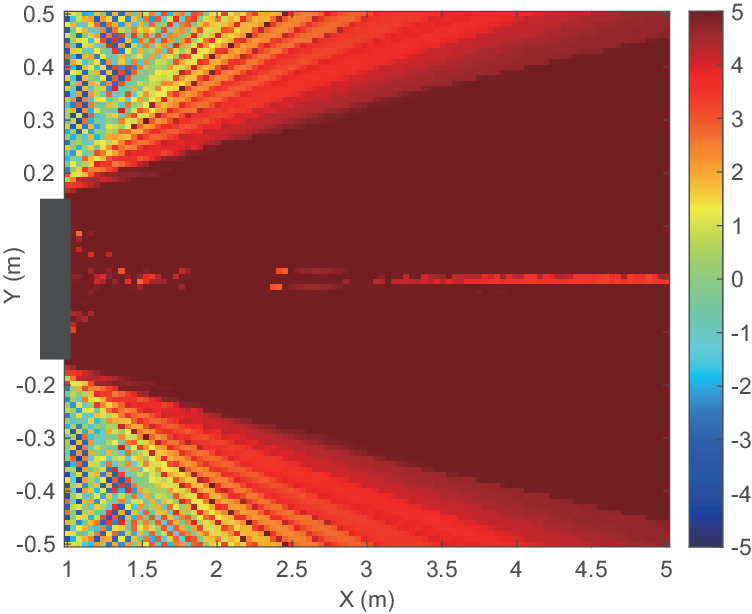}
	\caption{Gain improvement (dB) of hierarchical Airy beam training over exhaustive focused beam training.}
	\label{Fig:hie_exh}
\end{figure}

\section{Simulation Results}
\label{Sec-V}

In this section, we evaluate the performance of the proposed Airy beamforming enabled multi-user communication scheme in obstructed environments. The carrier frequency is set to $100$ GHz, and the BS is equipped with $N=266$ antenna elements. The number of RF chains is set to be equal to the number of users ($N_{\text{RF}} = K$). The obstacle is located at $x \in [0.8, 1]$ m and $y \in [-0.15, 0.15]$ m. For the near-field Airy beam codebook, the parameter space is sampled as follows: $\sin\theta$ is uniformly sampled with 90 points in $[\sin(-\pi/4), \sin(\pi/4)]$, $r$ is non-uniformly sampled with 6 points in $[1, 5]$ m, $a$ is uniformly sampled with 10 points in $[-2, 0]$, and $s$ is non-uniformly sampled with 20 points within $[-0.3, -0.05] \cup [0.05, 0.3]$.
The following baseline schemes are considered for comparison:
(1) \textbf{Perfect CSI}~\cite{hybrid}: Assuming perfect CSI is available at the BS, fully digital and hybrid beamformers are considered as performance upper bounds.
(2) \textbf{Focused Beamforming}~\cite{Multi1}: In the multiple access, the analog beamformer is designed using the near-field polar codebook. The sampling of the angle and distance domains is consistent with the near-field Airy beam codebook.
(3) \textbf{Steered Beamforming:}~\cite{SDMA}: The analog beamformer is designed based on the far-field DFT codebook, with angular sampling identical to that of the Airy beam codebook.

\begin{figure*}[t]
\setlength{\abovecaptionskip}{0pt}
\setlength{\belowcaptionskip}{0pt}
	\centering
    \includegraphics[width=1\linewidth]{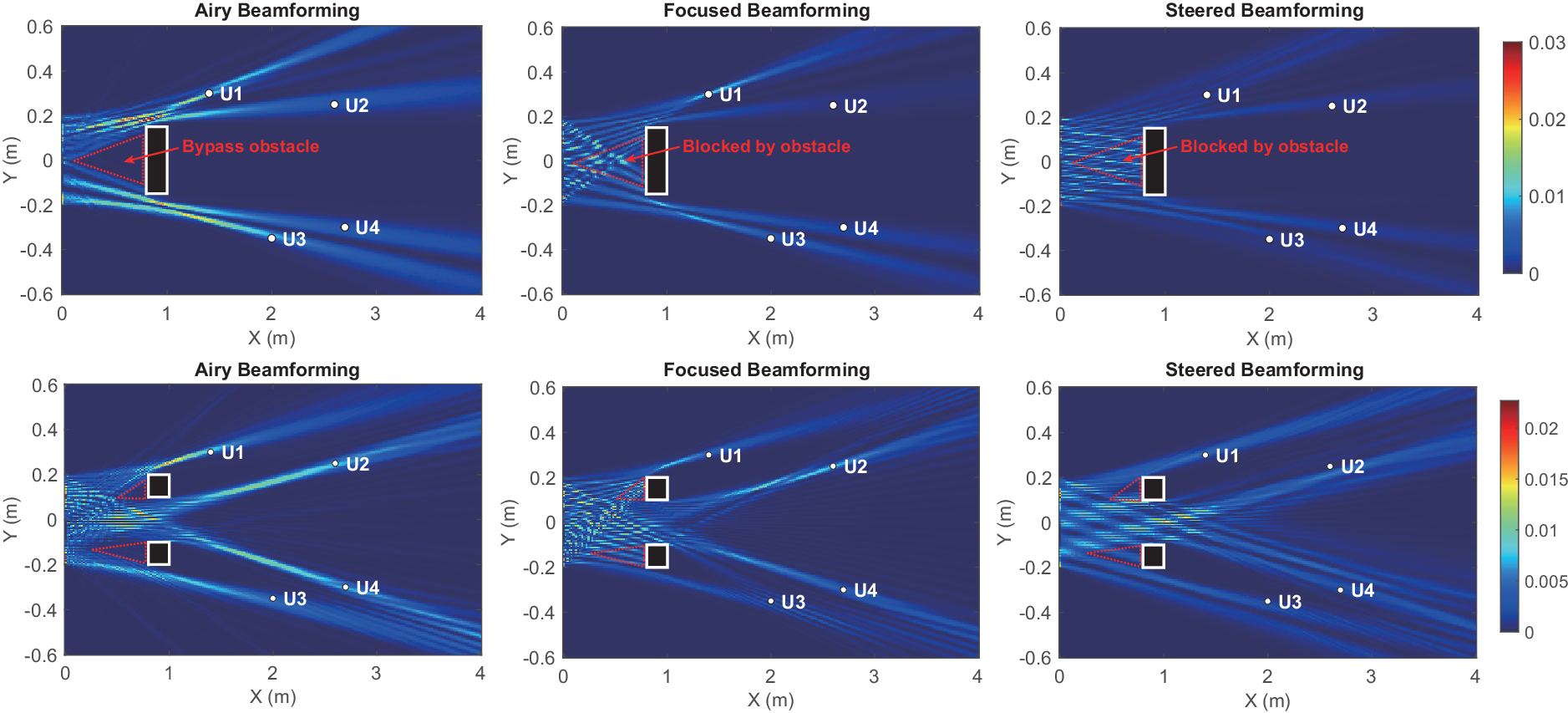}
	\caption{Beam patterns of various schemes after analog beamformer design with single and multiple obstacles.}
	\label{Fig:3patterns}
\end{figure*}

\begin{figure}[t]
\setlength{\abovecaptionskip}{0pt}
\setlength{\belowcaptionskip}{0pt}
	\centering
    \includegraphics[width=1\linewidth]{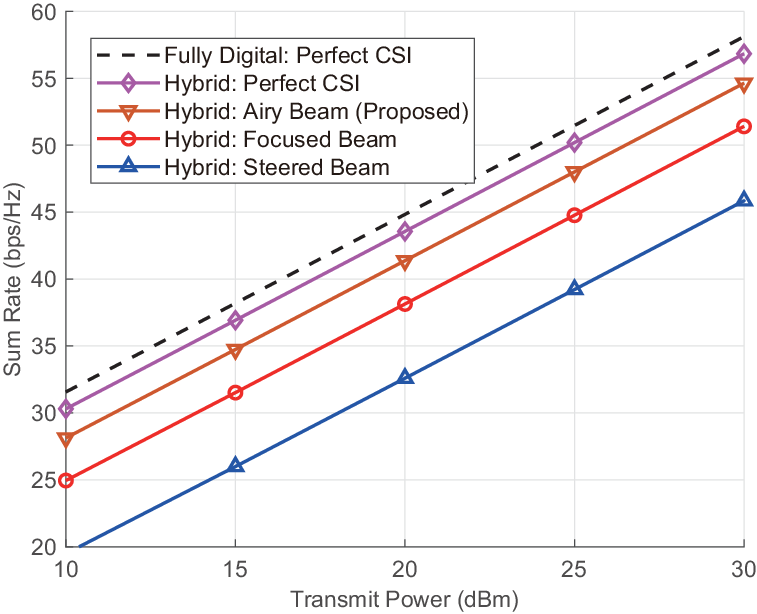}
	\caption{Sum rate vs. transmit power for a typical 4-user case.}
	\label{Fig:4rate}
\end{figure}

\subsection{Impact of Blockage Ratio on Received Power}

To evaluate the performance of the proposed near-field Airy beam in obstructed cases, Fig.~\ref{Fig:block_ratio} illustrates the received power versus the blockage ratio, defined as the proportion of obstructed LoS paths between the user and antenna elements. As the blockage ratio increases, the theoretical received power decreases rapidly, consistent with the performance upper bound discussed in Section~\ref{Sec_Guideline}.
When the blockage ratio equals 1, i.e., no LoS path exists between the user and any antenna element, the theoretical received power approaches zero, which validates~\emph{\textbf{Remark}}~\ref{remark1}.
At low blockage levels, the proposed Airy beam achieves performance comparable to the focused beam. Benefiting from the dual regulation of amplitude and phase responses, the Airy beam approaches the performance upper bound as blockage worsens, outperforming the focused beam. This demonstrates the versatility of the proposed wavefront in both obstructed and unobstructed cases. Moreover, due to the dominant near-field effects at close range, the far-field beam's performance degrades sharply.

\subsection{Performance of Hierarchical Airy Beam Training}

To evaluate the performance of the hierarchical Airy beam training scheme, we examine the spatial gain distribution across the service area. As shown in Fig.~\ref{Fig:exh_hie}, the received power achieved by the exhaustive search and the proposed three-stage hierarchical scheme using the same Airy codebook is compared. The results demonstrate that the hierarchical scheme achieves performance closely approaching the exhaustive search in most regions, with slight gaps occurring only in severely obstructed areas. This validates the effectiveness of the hierarchical search, which successfully reduces the training overhead from a multiplicative complexity to an additive one. Fig.~\ref{Fig:hie_exh} illustrates the gain improvement of hierarchical Airy beam training relative to exhaustive focused beam training. Benefiting from its enhanced match with obstructed channels, the Airy beam significantly outperforms the focused beam across most areas, particularly as blockage severity increases. Moreover, the training overhead of the hierarchical Airy search remains on the same level as that of the exhaustive focused beam search.

\begin{figure}[t]
\setlength{\abovecaptionskip}{0pt}
\setlength{\belowcaptionskip}{0pt}
	\centering
    \includegraphics[width=1\linewidth]{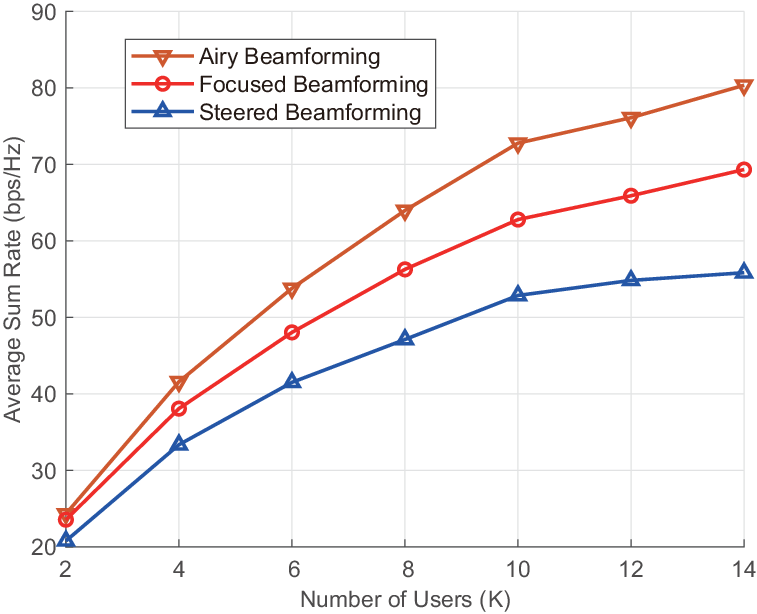}
	\caption{Sum rate vs. the number of users.}
	\label{Fig:user_num}
\end{figure}

\subsection{Beam Patterns of Multiple Access in Obstructed Environments}

Fig.~\ref{Fig:3patterns} illustrates a typical four-user obstructed case, where the LoS paths from the BS to the users are partially blocked. The beam patterns represent the performance of the hybrid beamformer after $\mathbf{F}_\text{A}$ is designed, reflecting the alignment of the codebook-based $\mathbf{F}_\text{A}$ with the target users. Benefiting from the match between the near-field Airy beam and the amplitude and phase responses of the obstructed channel, Airy beamforming can bypass obstacles and concentrate beam energy at the users in both single- and double-obstacle cases, thereby avoiding the spectral efficiency degradation caused by beam energy blockage. In contrast, for focused and steered beams, a majority of the energy fails to bypass the obstacles, leading to a degradation in received power.
Moreover, the optimal Airy beam for each user to bypass obstacles is closely related to the blockage condition. The proposed scheme adaptively selects the Airy beam that exploits the remaining LoS paths for propagation. As illustrated in Fig.~\ref{Fig:3patterns}, in the case of blockage by a single obstacle, the optimal Airy beam reaches the user by propagating along the edge of the obstacle, whereas in the double-obstacle case, the Airy beams can propagate through the gap between the two obstacles to reach the user.
Fig.~\ref{Fig:4rate} presents the sum rate of various schemes in the single-obstacle case. Although the introduction of the non-linear phase in the focused beam leads to an increase in the data rate, a gap still exists compared to the performance upper bound. The proposed Airy beamforming further enhances the data rate in multi-user obstructed environments and approaches the upper bound of perfect CSI.

\subsection{Impact of the Number of Users}

Fig.~\ref{Fig:user_num} depicts the sum rate performance of various multi-user transmission schemes with the number of users. The users are randomly distributed within a rectangular region defined by $x \in [1, 4]$ m and $y \in [-0.6, 0.6]$ m. Under a constant total transmit power constraint, it is observed that the sum rate increases with the number of users, whereas the average data rate per user gradually decreases due to power constraints and the escalation of multi-user interference. Across the entire range of user counts, the proposed Airy beamforming consistently outperforms both focused beamforming and steered beamforming. This is because, as the user count increases, Airy beams enable obstacle avoidance for multiple users, whereas focused and steered beams, which propagate toward a specific focal point and along a fixed direction, respectively, are prone to being directly intercepted by obstacles along their rectilinear trajectories. This demonstrates that in obstructed multi-user communications, Airy beamforming provides an effective solution for multi-user communications to circumvent obstacle-induced performance degradation.

\begin{figure}[t]
\setlength{\abovecaptionskip}{0pt}
\setlength{\belowcaptionskip}{0pt}
	\centering
    \includegraphics[width=0.95\linewidth]{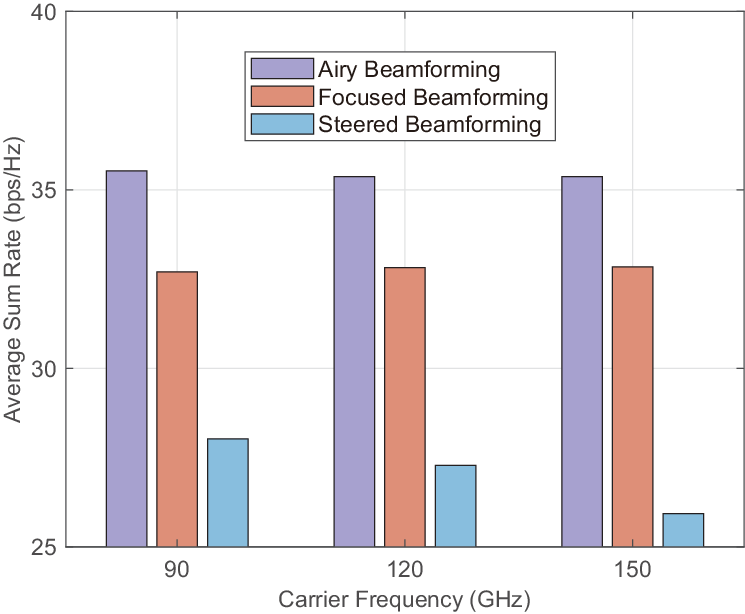}
	\caption{Sum rate vs. carrier frequency.}
	\label{Fig:fre}
\end{figure}

\subsection{Effect of the Carrier Frequency}

Fig.~\ref{Fig:fre} illustrates the variation of the sum rate versus frequency for four randomly distributed users. As the operating frequency increases toward the THz band, the performance of steered beamforming significantly degrades. This is attributed to the more pronounced near-field effects at higher frequencies, which lead to an inherent mismatch for the far-field steered beamformer. In contrast, Airy and focused beams account for the inherent non-linear phase in the near-field, thereby avoiding performance degradation as frequency increases. Moreover, the Airy beamforming enabled multi-user transmission scheme maintains favorable performance over focused beamforming across a wide frequency spectrum, ranging from typical mmWave bands (e.g., 90 GHz) to THz bands (e.g., 150 GHz). This confirms that in high-frequency obstructed near-field communications, Airy beams maintain effective obstacle avoidance and energy concentration to enhance data rates for multi-user access.

\begin{figure}[t]
\setlength{\abovecaptionskip}{0pt}
\setlength{\belowcaptionskip}{0pt}
	\centering
    \includegraphics[width=0.95\linewidth]{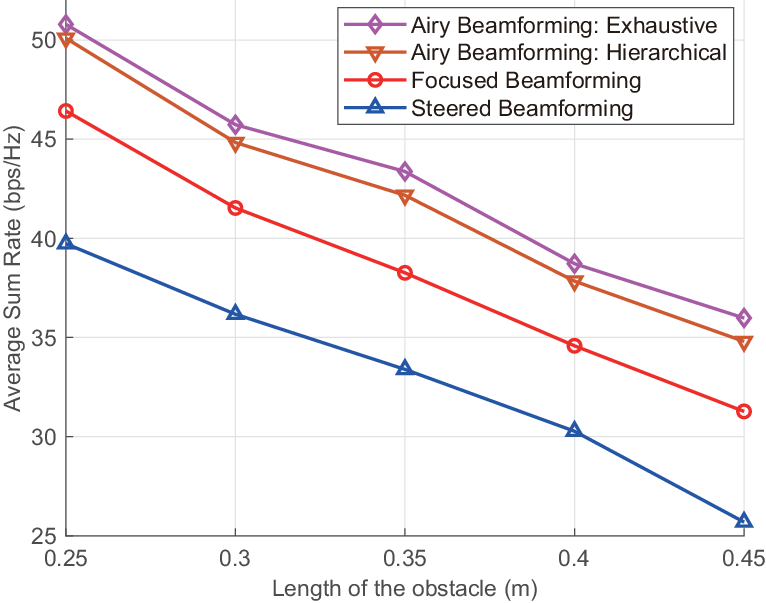}
	\caption{Sum rate vs. the length of the obstacle.}
	\label{Fig:length}
\end{figure}

\subsection{The Impact of the Obstacle Length}

Fig.~\ref{Fig:length} investigates the impact of obstacle dimensions on the sum rate, where the length of the obstacle is defined as its extent along the y-axis. With 4 users randomly distributed within the region $x \in [1, 5]$ m and $y \in [-1, 1]$ m, Fig.~\ref{Fig:length} shows that as the obstacle size increases, the region where LoS paths are obstructed expands. This leads to a drop in the users' received power and the average sum rate. Across various obstacle dimensions, the proposed Airy beamforming achieves a higher rate than focused beamforming due to its obstacle avoidance capability. Moreover, the performance of the proposed three-stage multi-user Airy beam training scheme closely approaches that of the exhaustive search. This validates that the scheme maintains high performance while significantly reducing training overhead from multiplicative to additive complexity. Compared to steered beamforming, focused beamforming exhibits a substantial performance gain by accounting for the near-field non-linear phase. However, a performance gap still exists compared to Airy beamforming because focused beams follow rectilinear trajectories and cannot bypass obstacles.

\section{Conclusions}
\label{Sec-VI}

In this paper, Airy beamforming enabled near-field multi-user communications are proposed to address the obstructed environments. To match the distinctive amplitude and phase characteristics of obstructed near-field channels, a new wavefront named near-field Airy beam is first introduced. Moreover, an Airy beam codebook and a hierarchical beam training scheme are proposed, which enable the selection of optimal Airy beams aligned with users at low overhead without relying on obstructed channel information. On this basis, an Airy beam based multi-user transmission scheme is developed for practical communications where the user and obstacle location are unknown. The Airy beam based hybrid beamformer design enables obstacle avoidance for multiple users to enhance spectral efficiency.

Simulation results demonstrate that: 1) The beam patterns demonstrate that the proposed hybrid Airy beamforming scheme simultaneously generates Airy beams for multiple near-field users, each bypassing obstacles while concentrating energy at the intended user.
2) The proposed Airy beamforming scheme adaptively accommodates varying obstacle locations and numbers, selecting the Airy beam propagation trajectory that exploits the LoS paths to bypass obstacles.
3) Under varying numbers of users, Airy beamforming further shifts the data rate toward the performance upper bound compared to conventional focused beamforming.

\end{document}